\documentclass[12pt]{iopart}
\usepackage{bm,color,bbm}

\usepackage{ulem} 
\expandafter\let\csname equation*\endcsname\relax
\expandafter\let\csname endequation*\endcsname\relax   

\usepackage{hyperref, mathtools,graphicx,amsmath,cases}

\newcommand{\beq}{\begin{equation}}
\newcommand{\eeq}{\end{equation}}
\newcommand{\bqa}{\begin{eqnarray}}
\newcommand{\eqa}{\end{eqnarray}}
\newcommand{\nn}{\nonumber}

\newcommand{\rt}[1]{\sqrt{#1}\,}

\newcommand{\smallfrac}[2]{\mbox{$\frac{#1}{#2}$}}

\newcommand{\half}{\smallfrac{1}{2}}

\newcommand{\id}{\mathbbm{1}}
\newtheorem{Theorem}{Theorem}

\newtheorem{Corollary}{Corollary}

\newcommand{\blk}{\color{black}}

\definecolor{maroon}{rgb}{0.7,0,0}
\definecolor{ngreen}{rgb}{0.3,0.7,0.3}

\definecolor{golden}{rgb}{0.8,0.6,0.1}

\newcommand{\B}{{\mathcal B}}
\newcommand{\str}{\mathcal{S}}
\newcommand{\R}{\mathcal R}

\usepackage{iopams}  
\begin{document}

\title{Generalising the Horodecki criterion to \blk nonprojective \blk qubit observables}

\author{Michael J. W. Hall$^1$ and Shuming Cheng$^{2,3,4}$}
\address{$^1$  Department of Theoretical Physics, Research School of Physics, Australian National University, Canberra ACT 0200, Australia}
\address{$^2$ The Department of Control Science and Engineering, Tongji University, Shanghai 201804, China}
\address{$^3$ Shanghai Institute of Intelligent Science and Technology, Tongji University, Shanghai 201804, China}
\address{$^4$ Institute for Advanced Study, Tongji University, Shanghai, 200092, China}

\vspace{10pt}

\begin{abstract}
The Horodecki criterion provides a necessary and sufficient condition for a two-qubit state to be able to manifest Bell nonlocality via violation of the Clauser-Horne-Shimony-Holt (CHSH) inequality. It requires, however, the assumption that suitable projective measurements can be made on each qubit, and is not sufficient for scenarios in which noisy or weak measurements are either desirable or unavoidable.  By characterising two-valued qubit observables in terms of strength, bias, and directional parameters, we address such scenarios by providing  necessary and sufficient conditions for arbitrary  qubit measurements having fixed strengths and relative angles for each observer. In particular, we find the achievable maximal values of the  CHSH parameter for unbiased measurements on arbitrary states, and, alternatively, for  arbitrary measurements on states with maximally-mixed marginals, and determine the optimal angles in some cases. We also show that for certain ranges of measurement strengths it is only possible to violate the CHSH inequality via biased measurements. \blk Finally, we use the CHSH inequality to obtain a simple necessary condition for the compatibility of two qubit observables. \blk
\end{abstract}

%
\vspace{2pc}
%
\submitto{\JPA}
%
\maketitle
%
%

\section{Introduction}

Bell nonlocality is a strikingly nonclassical feature of quantum correlations, which both strongly restricts possible interpretations of quantum phenomena~\cite{Bell64} and provides a useful physical resource for information tasks, such as secure quantum cryptography and randomness generation~\cite{Brunner14}.

The simplest and most common test for Bell nonlocality is violation of the Clauser-Horne-Shimony-Holt (CHSH) inequality~\cite{Clauser69}, 
\beq \label{chsh}
\blk S(X,X',Y,Y') \blk:=|\langle XY\rangle + \langle XY'\rangle+\langle X'Y\rangle - \langle X'Y'\rangle | \leq 2 ,
\eeq
by two space-like separated observers $A$ and $B$, where observer $A$ ($B$) measures one of two observables $X,X'$ ($Y,Y'$) on each run, with outcomes labeled by $\pm1$, and the angled brackets denote statistical averages. The quantity $\blk S(X,X',Y,Y') \blk$ is referred to as the CHSH parameter. The CHSH inequality can be violated by any sufficiently entangled quantum state, and by all pure entangled states in particular, via suitable choices of observables~\cite{Gisin91}. The degree of violation in a given setup determines corresponding bounds on the generation rate of secure cryptographic keys~\cite{Acin06}  and certified random numbers~\cite{Pironio10}.

The smallest quantum resource for Bell nonlocality, and frequently used in practical applications, is a two-qubit state shared by the observers. For such states the Horodecki criterion gives a necessary and sufficient condition for the existence of projective qubit observables $X$, $X'$, $Y$, and $Y'$ that violate the CHSH inequality~(\ref{chsh})~\cite{Horodecki95} (see also section~\ref{sec:horo}). In practice, however, the observers may not be able to measure projective  observables. This can be a consequence of, for example, apparatus limitations such as detector noise. But it also arises in contexts where it is necessary  to preserve some information or entanglement in the post-measurement state, as required by certain information protocols for entanglement recycling~\cite{Silva15}, randomness generation~\cite{Curchod17}, and state discrimination~\cite{Vargas21}.  In all such cases the Horodecki criterion is no longer a sufficient condition for the observers to be able to violate a Bell inequality. It is therefore of interest to generalise the criterion to nonprojective observables,  \blk i.e., generalised  observables, \blk under natural constraints. Several such generalisations are given in this paper.

In the following section, we briefly recap the description of \blk generalised \blk two-valued qubit observables, in terms of suitable strength, bias, and directional parameters. In section~\ref{sec:horo},  we generalise the Horodecki criterion to such observables. In particular, we provide necessary and sufficient conditions for violation of the CHSH inequality for  general states and unbiased observables in subsection~\ref{subsec3.1}, and for arbitrary observables and states with maximally-mixed marginals in subsection~\ref{subsec3.2}. We also obtain the optimal relative measurement angles for fixed strengths in several interesting cases in section~\ref{sec:optang}, \blk and use the CHSH inequality to find a simple necessary condition for the compatibility of qubit observables in section~\ref{sec:com}. \blk We conclude with a discussion of some implications of our results, and possible future work, in section~\ref{sec5}.\blk

\section{Two-valued qubit observables} \label{Sec2. Qunit Observables}

The outcomes of a general two-valued observable $X$ may be labeled by $\pm 1$ without  loss of generality, and the outcome statistics described by a positive-operator-valued-measure (POVM) $\{X_+,X_-\}$, with $X_\pm\geq0$ and $X_++X_-=\id$. The observable is projective if and only if $X_\pm$ are projection operators  (i.e., $X_\pm^2=X_\pm$). We follow the notation used in~\cite{Cheng21,ChengPRA} and represent the observable by the corresponding operator 
\beq \label{xop}
X:=X_+-X_-,\qquad -\id\leq X\leq \id .
\eeq
Thus, $X_\pm=\half(\id\pm X)$. The use of the same symbol for the observable and the operator is easily differentiated by context. For two such observables, $X$ and $Y$, measured on respective components of an entangled quantum state described by density operator $\rho$, one has the convenient result~\cite{ChengPRA}
\beq \label{prodxy}
\langle XY\rangle = \tr{\rho X\otimes Y} 
\eeq
for the quantities appearing in the CHSH inequality~(\ref{chsh}).  Thus, the average of the product of the observed outcomes is the average of the tensor product of the corresponding operators, even for nonprojective observables.

For qubit observables, the operator $X$ can be decomposed as
\beq 
X =  \B \id + \str\bm \sigma\cdot\bm x \label{observable}
\eeq
with respect to the Pauli spin operator basis $\boldsymbol{\sigma}\equiv (\sigma_1, \sigma_2, \sigma_3)$. Here $\B$ is the {\it bias} of the observable; $\str\geq0$ is its {\it strength} or {\it sharpness};  and $\bm x$  is a unit direction associated with the observable, with $|\bm x |:=(\bm x \cdot \bm x)^{1/2}=1$. 
It follows from equation~(\ref{xop}) that the strength and bias parameters satisfy the constraint
\beq \label{sbcon}
\str + |\B| \leq 1 .
\eeq
Hence, an observable with maximum strength $\str=1$ is unbiased, with $\B=0$, corresponding to the projective observable $X=\bm \sigma\cdot\bm x$ for spin in direction $\bm x$. In contrast, an observable with minimum strength $\str=0$ is trivial, with $X=\B\id$, corresponding to tossing a coin having biased outcome probabilities $\half(1\pm\B)$. The strength and bias parameters also determine the maximum reversibility of measurements of $X$~\cite{Cheng21,ChengPRA}.

A useful form of the expectation value $\langle XY\rangle$ in equation~(\ref{prodxy}) can be found via the Fano form of the two-qubit density operator $\rho$, i.e.,
\beq \label{bloch}
\rho=\frac14\sum_{\mu,\nu=0}^4 \Theta_{\mu\nu} \, \sigma_\mu\otimes\sigma_\nu,\qquad \Theta:=\begin{pmatrix} 1 & \bm b^\top\\ \bm a & T \end{pmatrix} .
\eeq
Here $\sigma_0=\id$; $\bm a:=\langle \bm \sigma\otimes \id\rangle$ and $\bm b:=\langle \id\otimes \bm \sigma\rangle$ are the qubit Bloch vectors for observers $A$ and $B$; and $T:=\langle \bm\sigma\otimes\bm\sigma^\top\rangle$ is the spin correlation matrix. Substitution of equations~(\ref{observable}) and~(\ref{bloch}) into equation~(\ref{prodxy}) then gives
\beq \label{prodxy2}
\langle XY\rangle = \begin{pmatrix} \B_X&  \str_X \bm x^\top\end{pmatrix}
\begin{pmatrix} 1 & \bm b^\top\\ \bm a & T \end{pmatrix} 
\begin{pmatrix} \B_Y\\ \str_Y \bm y \end{pmatrix} ,
\eeq
for $X=\B_X\id+\str_X\bm\sigma\cdot\bm x$ and $Y=\B_Y\id+\str_Y\bm\sigma\cdot\bm y$. This expression is a key ingredient for the results of the next section.

\section{Generalising the Horodecki criterion}
\label{sec:horo}

\subsection{The Horodecki criterion}\label{Horodeckicriterion}

The Horodecki criterion gives a necessary and sufficient condition for the existence of qubit observables $X$, $X'$, $Y$, and $Y'$ that violate the CHSH inequality~(\ref{chsh}), for a given two-qubit state~\cite{Horodecki95}:
\beq \label{horo}
H(T)  :=  2\sqrt{s_1(T)^2 + s_2(T)^2} > 2 . 
\eeq
Here $T$ is the $3\times3$ spin-correlation matrix in equation~(\ref{bloch}), and $s_1(A), s_2(A),\dots$ denote the singular values of \blk any given \blk matrix $A$ (i.e., the square roots of the eigenvalues of $A^\top A$), in decreasing order. 

Note that the Horodecki parameter $H(T)$ is not an upper bound for the CHSH parameter in general. For example, for  $X=X'=Y=Y'=\B\id$ and $T=0$ one has $\blk S(X,X',Y,Y') \blk=2\B^2\geq H(T)=0$. However, $H(T)$ does represent the maximum possible value of the CHSH parameter for the case of projective observables with unit strengths~\cite{Horodecki95}, \blk and indeed for all unbiased observables~\cite{Loubenets20} (see also section~\ref{subsec3.1}). \blk  Moreover, the criterion $H(T)>2$ for violation of the CHSH inequality remains  valid for  arbitrary  observables~\cite{Cheng21,Cleve04}.  Indeed, it follows from the general convexity argument in~\cite{Cleve04}, or more directly via the proof of equation~(S31) in~\cite{Cheng21}, that one has the tight upper bound
\beq \label{supper}
\blk S(X,X',Y,Y') \blk \leq \max \{ 2, H(T) \} ,
\eeq
where the upper bound is always attainable (noting that the choice of zero-strength projective observables $X=X'=Y=Y'=\id$ trivially yields $\blk S(X,X',Y,Y') \blk=2$). 

However, the observers  may not be able to measure unit-strength  observables in some scenarios---e.g., as noted in the introduction, due to detector noise or to a need to preserve some information or entanglement in the post-measurement state~\cite{Silva15,Curchod17,Vargas21}. In such cases the bound in equation (\ref{supper}) is no longer always achievable,  and the Horodecki criterion is no longer a sufficient condition for being able to violate a Bell inequality. It is therefore of interest to generalise the criterion, under natural constraints on strengths, biases, and/or measurement directions. Several such generalisations are given in the remainder of this paper.

\subsection{Generalised criterion for unbiased observables}\label{subsec3.1}

We first consider the case where the observers make measurements of unbiased observables, corresponding to $\B=0$ in equation~(\ref{observable}). This case applies whenever their detectors respond to the maximally random input state $\rho=\half\id$ with a maximally random output distribution $p_\pm=\half$, which is often applicable in noisy and weak measurement scenarios, as well as in the context of unbiased estimates of observables. For example, most previous work on the recycling of Bell nonlocality via suitably weak measurements has been confined to unbiased observables~\cite{Silva15,Curchod17,Mal16,Bera18,Shenoy19,Brown20,Maity20,Zhang21,Schiavon17,Hu18,Choi20,Feng20,Jie21}. 

Letting $\str_X, \str_{X'}, \str_Y, \str_{Y'}$ and $\bm x, \bm x', \bm y, \bm y'$ denote the respective strengths and measurement directions of unbiased qubit observables $X,X',Y,Y'$, the CHSH parameter in equation~(\ref{chsh}) simplifies via Eq.~(\ref{prodxy2}) to
\begin{align} \label{chshzero}
	\blk S(X,X',Y,Y') \blk &= |\str_X\str_{Y} \bm x^\top T\bm y + \str_X\str_{Y'} \bm x^\top T\bm y'+ \str_{X'}\str_{Y} \bm x'^\top T\bm y - \str_{X'}\str_{Y'} \bm x'^\top T\bm y'| .
\end{align}
We then have the following generalisation of the Horodecki criterion, proved in~\ref{appa1}.
\begin{Theorem} \label{theorem1} For unbiased observables $X,X',Y,Y'$, with fixed strengths $\str_X, \str_{X'}, \str_Y,\str_{Y'}$ and relative angles $\cos\theta=\bm x\cdot\bm x'$,  $\cos\phi=\bm y\cdot\bm y'$, measured on a two-qubit state with correlation matrix $T$, the CHSH parameter has the tight upper bound
	\begin{align} \label{sabzero}
	\blk S(X,X',Y,Y') \blk\leq S_0 &:=s_1(T)s_1(W)+s_2(T) s_2(W) \\ 
	&=\half  [s_1(T)+s_2(T)]I_+(W) +  \half  [s_1(T)- s_2(T)]I_-(W) ,
	\label{cor2}
	\end{align}
where $W$ is the $2\times2$ matrix
	\beq \label{wdef}
	W:=\begin{pmatrix}
		A\cos\frac{\theta}{2}\cos\frac{\phi}{2} & B\cos\frac{\theta}{2}\sin\frac{\phi}{2} \\ C\sin\frac{\theta}{2}\cos\frac{\phi}{2} & -D\sin\frac{\theta}{2}\sin\frac{\phi}{2} 
	\end{pmatrix}
	\eeq
	with
\begin{align}
	A &= \str_{X}\str_{Y}+\str_{X}\str_{Y'}+\str_{X'}\str_{Y}-\str_{X'}\str_{Y'}\nn\\
	B &= \str_{X}\str_{Y}-\str_{X}\str_{Y'}+\str_{X'}\str_{Y}+\str_{X'}\str_{Y'} \nn\\
	C &= \str_{X}\str_{Y}+\str_{X}\str_{Y'}-\str_{X'}\str_{Y}+\str_{X'}\str_{Y'} \nn\\
	D &= -\str_{X}\str_{Y}+\str_{X}\str_{Y'}+\str_{X'}\str_{Y}+\str_{X'}\str_{Y'} ,
	\label{abcd} 
\end{align}
and where $I_\pm(W):=s_1(W)\pm s_2(W)\geq0$ may be calculated explicitly via
\begin{align} I_\pm(W)^2&=
	(\str_{X}^2+\str_{X'}^2)(\str_{Y}^2+\str_{Y'}^2) +2\str_X\str_{X'}(\str_{Y}^2-\str_{Y'}^2)\cos\theta \nn\\ &~~+2\str_Y\str_{Y'}(\str_{X}^2-\str_{X'}^2)\cos\phi 
\pm 4\str_{X}\str_{X'}\str_{Y}\str_{Y'} \sin\theta\sin\phi .
	\label{iw}
\end{align}
\end{Theorem}

It immediately follows from this theorem that the CHSH inequality can be violated by measurements of unbiased observables, \blk with given strengths and relative angles, \blk if and only if $S_0>2$. Further, for the special case of a singlet state one has $s_1(T)=s_2(T)=1$ and the upper bound reduces to $S_0=I_+(W)$, recently obtained (by less general means) in~\cite{ChengPRA}, where it was a key ingredient for obtaining one-sided monogamy relations for qubit recycling. Theorem~\ref{theorem1} represents a substantial generalisation of this special case, valid for all two-qubit states. 
Note that the upper bound is, like Bell nonlocality itself, invariant under local unitary transformations on each side (since such transformations preserve measurement strengths, the relative angles $\theta$ and $\phi$, and the singular values of the spin correlation matrix). As discussed in~\ref{appa}, the optimal measurement directions for given values of 
 $\theta$ and $\phi$ may be obtained by applying the orthogonal transformations in equation~(\ref{o1o2}).

A simple link between Theorem~1 and the Horodecki criterion~(\ref{horo}) is evident from the upper bound
\beq \label{schwarz}
\blk S(X,X',Y,Y') \blk^2\leq \tr{W^\top W} \left[s_1(T)^2+s_2(T)^2\right]\leq H(T)^2 
\eeq 
for unbiased observables, where the first inequality follows from equation~(\ref{sabzero}) via the Schwartz inequality and the identity $\tr{W^\top W}=\sum_j s_j(W)^2$, and the second via $\tr{W^\top W}\leq \max\{A^2,B^2,C^2,D^2\}\leq 4$. Thus, the Horodecki parameter $H(T)$ is an upper bound for the CHSH parameter for the case of unbiased observables, \blk as also follows from Theorem~2 in~\cite{Loubenets20} (noting that $X$ in (\ref{observable}) is traceless when $\B=0$). \blk

A closer link with the Horodecki criterion is given by the following corollary. 
\begin{Corollary} \label{corollary1}
 For unbiased observables $X,X',Y,Y'$ with equal strengths on each side, i.e., $\str_X=\str_{X'}$ and $\str_Y=\str_{Y'}$,  measured on a two-qubit state with correlation matrix $T$, the CHSH parameter has the tight upper bound
	\beq \label{cor1}
	\blk S(X,X',Y,Y') \blk \leq 2\str_X \str_Y\sqrt{s_1(T)^2+s_2(T)^2}  ,
	\eeq
with the bound achievable for any relative angles satisfying
\beq \label{optang}
\sin \theta \sin \phi = \frac{2s_1(T)s_2(T)}{s_1(T)^2+s_2(T)^2}  .
\eeq
\end{Corollary}

This corollary is derived in~\ref{appa2}, and its tightness immediately implies the necessity and sufficiency of the Horodecki criterion  in equation~(\ref{horo}) for the case of projective observables with unit strengths (i.e., $\str_X=\str_{X'}=\str_Y=\str_{Y'}=1$).  The assumption of equal strengths on each side is reasonable in scenarios such as the measurement of linear photon polarisations via a rotatable polariser (although less so in intrinsically anisotropic scenarios such as two-level atoms, where the measurement direction corresponding to the energy basis is different in kind to other directions). Note that for states with $s_1(T)=s_2(T)$  (e.g, maximally entangled states and Werner states), equation~(\ref{optang}) requires orthogonal relative angles, $\theta=\phi=\pi/2$, to achieve the upper bound. More generally a range of optimal angles is possible, but always includes the choice of equal angles specified by 
\beq \label{eqang} 
\sin\theta=\sin\phi=[2s_1(T)s_2(T)]^{1/2}/[s_1(T)^2+s_2(T)^2]^{1/2}\leq1. 
\eeq 
It would be of interest to determine the optimal relative angles for any given set of strengths, and some partial results in this regard are given in section~\ref{sec:optang} below, including a significant generalisation of Corollary~\ref{corollary1}.

Since the bound in Theorem~\ref{theorem1} is always achievable, it follows that sufficient conditions for being able to violate the CHSH inequality, that depend only on the strengths of the observables, can be obtained by choosing specific values of the relative angles $\theta$ and $\phi$. For example, choosing orthogonal angles $\theta=\phi=\pi/2$ and substituting into equations~(\ref{cor2}) and~(\ref{iw}) yields the following sufficient condition.
\begin{Corollary} \label{corollary2} For unbiased observables $X,X',Y,Y'$ with fixed strengths $\str_X, \str_{X'}, \str_Y,\str_{Y'}$, a sufficient condition for being able to violate the CHSH inequality on a two-qubit state with spin correlation matrix $T$ is 
	\beq \label{cor3}
	S_0^\perp :=\half  (i_++i_-)s_1(T) + \half  (i_+-i_-)s_2(T) > 2,
	\eeq
	where
\beq \label{vpm}
i_\pm:= \sqrt{(\str_X\str_Y\pm \str_{X'}\str_{Y'})^2 + (\str_X\str_{Y'}\pm \str_{X'}\str_{Y})^2}.
\eeq
\end{Corollary}
Equation~(\ref{cor3}) is also a necessary condition for the case of equal strengths on each side and equal singular values $s_1(T)=s_2(T)$, as per  the discussion following Corollary~\ref{corollary1},  i.e., orthogonal relative angles are optimal for this case.

Finally, note that $I_\pm(W)$ in equation~(\ref{iw}), and hence the bound $S_0$ in Theorem~\ref{theorem1}, are not invariant under the interchange of $X$ and $X'$ or of $Y$ and $Y'$ in general, due to presence of terms depending on the differences of the corresponding strengths. This reflects the fact that the CHSH inequality in equation~(\ref{chsh}) itself is not invariant under such interchanges, but transforms between four different versions~\cite{Brunner14}. Since the relative angles between the measurement directions on each side are not affected by these  interchanges, and $\sin\theta,\sin\phi\geq0$, this directly leads to a necessary and sufficient condition for being able to violate any one of the  four CHSH inequalities as follows.
\begin{Corollary} \label{corollary3} For unbiased observables $X,X',Y,Y'$, with fixed strengths $\str_X, \str_{X'}, \str_Y,\str_{Y'}$ and relative angles $\cos\theta=\bm x\cdot\bm x'$,  $\cos\phi=\bm y\cdot\bm y'$, measured on a two-qubit state with correlation matrix $T$, one of the four possible CHSH inequalities can be violated if and only if
\begin{align} \label{tildeszero}
\tilde S_0:=\half  [s_1(T)+s_2(T)]\tilde I_+(W) +  \half  [s_1(T)- s_2(T)]\tilde I_-(W) >2,
\end{align}
with $\tilde I_\pm(W) \geq 0$ defined via 
\begin{align} \tilde I_\pm(W)^2&=
	(\str_{X}^2+\str_{X'}^2)(\str_{Y}^2+\str_{Y'}^2) +2\str_X\str_{X'}|\str_{Y}^2-\str_{Y'}^2|\,|\cos\theta| \nn\\
	& +2\str_Y\str_{Y'}|\str_{X}^2-\str_{X'}^2|\,|\cos\phi |
	 \pm 4\str_{X}\str_{X'}\str_{Y}\str_{Y'} \sin\theta\sin\phi .
	\label{tildeiw}
\end{align}
\end{Corollary}
Note for the case of equal strengths on each side, $\str_X=\str_{X'}$ and $\str_Y=\str_{Y'}$, one has $\tilde S_0=S_0$, and hence the condition $S_0>2$ is necessary and sufficient for this case, i.e, one of the CHSH inequalities can be violated if and only if the canonical CHSH inequality~(\ref{chsh}) can be violated.

\subsection{Generalised criterion for T-states}\label{subsec3.2}

We will now obtain a generalised Horodecki criterion that is valid for {\it all} observables, whether biased or unbiased, under the condition that the local qubit states are maximally mixed. This condition corresponds to  $\bm a=\bm b=\bm 0$ in equation~(\ref{bloch}), yielding $\rho=\frac14(\id\otimes \id +\bm \sigma^\top\otimes T\bm \sigma)$. Such states  are therefore fully characterised by their spin correlation matrix $T$,  and are commonly referred to T-states~\cite{Horodecki96}. They include maximally entangled states and Werner states in particular.

The form of the CHSH parameter for T-states follows via equations~(\ref{chsh}) and~(\ref{prodxy2}) as 
\begin{align}
	\blk S(X,X',Y,Y') \blk &= |\str_X\str_{Y} \bm x^\top T\bm y + \str_X\str_{Y'} \bm x^\top T\bm y' + \str_{X'}\str_{Y} \bm x'^\top T\bm y - \str_{X'}\str_{Y'} \bm x'^\top T\bm y' + J| ,
	\label{sabt}
\end{align}
with
\beq \label{jay}
J:=\B_{X}\B_{Y}+\B_{X}\B_{Y'}+\B_{X'}\B_{Y}-\B_{X'}\B_{Y'}.
\eeq
Note that this form only differs from that in equation~({\ref{chshzero}) for unbiased observables by the additional term $J$. This leads to a relatively straightforward analogue of Theorem~\ref{theorem1}, as follows.
\begin{Theorem} \label{theorem2}  For arbitrary observables $X,X',Y,Y'$ with fixed strengths $\str_X, \str_{X'}, \str_Y,\str_{Y'}$ and relative angles $\cos\theta=\bm x\cdot\bm x'$,  $\cos\phi=\bm y\cdot\bm y'$,  measured on a T-state with correlation matrix $T$, the CHSH parameter has the tight upper bound
		\begin{align} \label{thm2}
			\blk S(X,X',Y,Y') \blk\leq S_T &:= S_0 + J_{\max} ,
		\end{align}
		where $S_0$ is given in equation~(\ref{sabzero}) and
	\begin{align}
		J_{\max}	&:= (2-\str_X-\str_{X'})(2-\str_Y-\str_{Y'}) - 2(1-\max\{\str_X,\str_{X'}\})(1-\max\{\str_Y,\str_{Y'}\}) .
		\label{jmax}
	\end{align}
\end{Theorem}
	
	It follows immediately that the CHSH inequality~(\ref{chsh}) can be violated for \blk given strengths and relative angles on \blk T-states  if and only if $S_T\geq 2$. Theorem~\ref{theorem2} is proved in~\ref{appa1}, where it is shown that the form of $J_{\max}$ arises from optimising the choice of biases subject to the constraint $\str+|\B|\leq1$ in equation~(\ref{sbcon}). 
	
	One has $J_{\max}\geq0$ from the second line of equation~(\ref{proofthm2}), and hence, comparing Theorems~\ref{theorem1} and~\ref{theorem2}, biased observables can typically achieve a greater value of the CHSH parameter for T-states than can unbiased observables. An important exception is  the case of projective observables with unit strengths, for which $J_{\max}=0$ and $S_T=S_0$, implying that unbiased observables are optimal for this case.
	
	Further, $J_{\max}\leq 2$ from the second line of equation~(\ref{proofthm2}), with equality for the case of zero strengths. However, for this case $I_\pm(W)$ in equation~(\ref{iw}) vanishes, so that $S_0=0$ and $S_T=2$, implying the CHSH inequality cannot be violated.  This raises an interesting question: are there cases where the CHSH inequality can be violated by measuring biased  observables, but not by measuring unbiased observables? We answer this question in the affirmative in Corollary~\ref{corollary5}, further below. 
	
	The relation $S_T=S_0+J_{\max}$ between the upper bounds in Theorems~\ref{theorem1} and~\ref{theorem2} immediately leads to analogues of Corollaries~\ref{corollary1}--\ref{corollary3}  for the measurement of general observables on T-states. For example, as shown in~\ref{appa2}, the analogue of Corollary~\ref{corollary1} is given by the following.
	\begin{Corollary} \label{corollary4} For arbitrary observables $X,X',Y,Y'$ with equal strengths on each side, i.e., $\str_X=\str_{X'}$ and $\str_Y=\str_{Y'}$, measured on a T-state, the CHSH parameter has the tight upper bound
		\begin{align} \label{cor4}
			\blk S(X,X',Y,Y') \blk &\leq 2\str_X \str_Y \sqrt{s_1(T)^2+s_2(T)^2} +2(1-\str_X)(1-\str_Y)  ,
		\end{align}
	where the bound is achievable for relative angles satisfying
	\beq \label{optangt}
	\sin \theta \sin \phi = \frac{2s_1(T)s_2(T)}{s_1(T)^2+s_2(T)^2}  .
	\eeq
	\end{Corollary}
	Thus, similarly to the case of unbiased observables in Corollary~\ref{corollary1}, orthogonal relative angles, $\theta=\phi=\pi/2$, are optimal for arbitrary observables measured on T-states if $s_1(T)=s_2(T)$. 
	
	This result puts us in a position to answer the question posed above, as to whether it is optimal to measure biased rather than unbiased observables in some cases. In particular, consider a simple example in which each observable has the same strength $\str$ (corresponding, for example, to isotropic detector noise), and the observers share a T-state. Then from Corollary~\ref{corollary1} the maximum achievable violation of the CHSH inequality via measurement of unbiased observables is
	\beq \label{unbiased}
	\blk S(X,X',Y,Y') \blk_{\rm unbiased} = 2\str^2 \sqrt{s_1(T)^2+s_2(T)^2} ,
	\eeq
	whereas from Corollary~\ref{corollary4} the maximum achievable violation via measurement of biased observables is
	\beq \label{biased}
		\blk S(X,X',Y,Y') \blk_{\rm biased} = 2\str^2 \sqrt{s_1(T)^2+s_2(T)^2} + 2(1-\str)^2 .
	\eeq
	It follows that if the CHSH inequality can be violated by unbiased observables for a given value of $\str<1$, then it can be violated by an even greater amount for biased observables with the same strength. Further, there are values of $\str$ for which the CHSH inequality can {\it only} be violated by biased observables. In particular, writing $R=\sqrt{s_1(T)^2+s_2(T)^2}$, then equations~(\ref{unbiased}) and~(\ref{biased}) imply that a violation $\blk S(X,X',Y,Y') \blk>2$  requires 
	\beq
	\str>\str_{\rm unbiased}:=1/\sqrt{R},\qquad \str>\str_{\rm biased}:=2/(1+R),
	\eeq
	for unbiased and biased observables, respectively. Hence, for any strength satisfying $\str_{\rm unbiased} \geq \str > \str_{\rm biased}$ (which is always possible if $R>1$, i.e., if $H(T)>2$), the CHSH inequality can be violated by biased observables but not by unbiased observables.  
	
	We encapsulate the above observations  in the following Corollary.
	\begin{Corollary} \label{corollary5} 
	The maximum possible value of the CHSH parameter, for observables with a fixed set of strengths $\str_X,\str_{X'},\str_Y,
	\str_{Y'}$, can be strictly greater for biased observables than for unbiased observables. Moreover, there are cases where the CHSH inequality can be only violated by biased observables.  
	\end{Corollary}
	
	Finally,  we note that the T-state analogue of Corollary~\ref{corollary2} is that 
	\beq
	S_0^\perp+J_{\max} >2
	\eeq
	is a sufficient condition to be able to violate the CHSH inequality via general measurements of given strengths on a T-state, where $S_0^{\perp}$ is defined in equation~(\ref{cor3}), while the analogue of Corollary~\ref{corollary3} is given by the following.
	\begin{Corollary} \label{corollary6} For arbitrary observables $X,X',Y,Y'$, with fixed strengths $\str_X, \str_{X'}, \str_Y,\str_{Y'}$ and relative angles $\cos\theta=\bm x\cdot\bm x'$,  $\cos\phi=\bm y\cdot\bm y'$, measured on a T-state, one of the \blk four \blk possible CHSH inequalities can be violated if and only if
		\begin{align} \label{tildest}
			\tilde S_T:=\tilde S_0 + J_{\max} > 2,
		\end{align}
		where $\tilde S_0$ is defined in equation~(\ref{tildeszero}).
	\end{Corollary}
Note for the case of equal strengths on each side, $\str_X=\str_{X'}$ and $\str_Y=\str_{Y'}$, one has $\tilde S_T=S_T$. Hence the condition $S_T>2$ is a necessary and sufficient condition for this case.

\subsection{A general necessary criterion}

We close this section by noting a formal approach that gives a necessary condition for violation of the CHSH inequality for arbitrary observables and states, and which may provide the basis for obtaining more general results in future work. 

In particular, note from equations~(\ref{chsh}) and~(\ref{prodxy2}) that the CHSH parameter may be rewritten for general qubit observables and two-qubit states in the form
\beq \label{sabgen}
\blk S(X,X',Y,Y') \blk = |\tr{\Theta N^\top}|,
\eeq
where  $\Theta$ is the $4\times4$ matrix in equation~(\ref{bloch}) and $N$ is the $4\times4$ matrix given by $N:=\bm u_X \bm u_Y^\top + \bm u_X \bm u_{Y'}^\top + \bm u_{X'} \bm u_Y^\top -\bm u_{X'} \bm u_{Y'}^\top$, where $\bm u_X$ denotes the 4-vector $(\B_X,\str_X \bm x)^\top$ for observable $X$. The proof of equation~(\ref{lemmaproof}) in~\ref{appa1}  then easily generalises to give the general upper bound
\beq \label{sgen}
\blk S(X,X',Y,Y') \blk \leq S_{\rm gen} := \sum_j s_j(\Theta) s_j(N).
\eeq
for the CHSH parameter. This immediately yields a strong necessary condition, $S_{\rm gen}>2$,  for violation of the CHSH inequality for arbitrary given measurements on an arbitrary given two-qubit state.

\section{Optimal angles for fixed strengths}
\label{sec:optang}

The tight upper bounds in Theorems~\ref{theorem1} and~\ref{theorem2} are functions of both the measurement strengths and the relative angles on each side, implying the desirable feature of invariance under local rotations (like Bell nonlocality itself). The dependence on relative angles is valuable in some cases --- for example, it is an essential ingredient for obtaining one-sided monogamy relations that restrict the sharing of Bell nonlocality via qubit recycling~\cite{Cheng21,ChengPRA}.  However, noting that it is often more straightforward to control measurement directions (e.g., via rotation of a polariser) than measurement strengths (which may be set by apparatus limitations or environmental noise), it is of interest to determine the optimal relative angles that maximise the CHSH parameter for a fixed set of measurement strengths. This can, of  course, always be done by numerically optimising $S_0$ in equations~(\ref{cor2}) and~(\ref{thm2}), over $\theta$ and $\phi$. Analytic results are more illuminating, however, and we give several such results here.

We first recall that the optimal relative angles have already been determined for the special case of equal strengths on each side, in Corollaries~\ref{corollary1} and~\ref{corollary4}. In this case the optimal angles are degenerate, i.e., any values of $\theta$ and $\phi$ satisfying~(\ref{optang}), such as those in equation~(\ref{eqang}), are optimal. However, we show that this degeneracy is lifted more generally, including for the case of equal strengths on just one side (Theorem~\ref{theorem3}). We also determine the optimal relative angles for the case of arbitrary strengths and equal singular values $s_1(T)=s_2(T)$ of the spin correlation matrix (Theorem~\ref{theorem4}), and show violation of the CHSH inequality remains possible in the limit that one strength becomes arbitrarily small.

Our first result is a generalisation of Corollaries~\ref{corollary1} and~\ref{corollary4}.
\begin{Theorem} \label{theorem3}
For observables $X,X',Y,Y'$ with equal strengths on one side, measured on a two-qubit state with correlation matrix $T$,  and taking $\str_X=\str_{X'}$ and $\str_Y\geq \str_{Y'}$ without loss of generality, the CHSH parameter has the tight upper bounds
\begin{align} \label{thm30}
\blk S(X,X',Y,Y') \blk&\leq  2\str_X \sqrt{ s_1(T)^2 \str_Y^2 + s_2(T)^2 \str_{Y'}^2} , \\
 \label{thm3t}  
\blk S(X,X',Y,Y') \blk&\leq   2\str_X \sqrt{ s_1(T)^2 \str_Y^2 + s_2(T)^2 \str_{Y'}^2}+2(1-\str_X)(1-\str_{Y'}),
\end{align}
for the respective cases of unbiased measurements on arbitrary states and arbitrary measurements on T-states. Further, these bounds are achieved for the relative angles given by
\beq \label{optangone}
 \sin \theta = \frac{2s_1(T)s_2(T)\str_Y\str_{Y'}}{ s_1(T)^2 \str_Y^2 + s_2(T)^2 \str_{Y'}^2}  , \qquad
\phi=\frac{\pi}{2}.
\eeq
\end{Theorem}

The above result is proved in~\ref{appc}, and the upper bounds represent substantial generalisations of the Horodecki parameter~(\ref{horo}). Note that in the limit of equal strengths on {\it both} sides, i.e., if one also has $\str_Y=\str{_Y'}$, then the upper bounds reduce to those in Corollaries~(\ref{corollary1}) and~(\ref{corollary4}) for the respective cases, as expected. Further, the optimal angles in equation~(\ref{optangone}) satisfy equation~(\ref{optang}). However, a  notable difference is that the optimal angles are specified uniquely for $\str_Y\neq\str_{Y'}$, as per equation~(\ref{optangone}), whereas there is a degenerate one-parameter family of optimal angles for $\str_Y=\str_{Y'}$, as per equation~(\ref{optang}).  In either case, one has the interesting property that if one observer measures observables with equal strengths, then  orthogonal measurement directions are an optimal choice for the other observer.

The optimal angles can also be determined if the first two singular values of the spin correlation matrix are equal, i.e., $s_1(T)=s_2(T)$, as per the following result.
\begin{Theorem} \label{theorem4}
For observables $X,X',Y,Y'$, measured on a two-qubit state with spin correlation matrix $T$ having equal singular values $s_1(T)=s_2(T)$,  the CHSH parameter has the tight upper bounds
\begin{align} \label{thm4upper}
	\blk S(X,X',Y,Y') \blk \leq  S^*_0 , \qquad	\blk S(X,X',Y,Y') \blk \leq S^*_0 + J_{\max} ,
\end{align}
for the respective cases of unbiased measurements on arbitrary states and arbitrary measurements on T-states, where 
\begin{align} \label{thm4}
S^*_0 := \left\{ 
\begin{array}{ll}
s_1(T) \sqrt{2(\str_{X}^2+\str_{X'}^2) (\str_{Y}^2+\str_{Y'}^2)}	 , & \qquad  \frac{|\str_X^2-\str_{X'}^2|}{2\str_X\str_{X'}}  \frac{|\str_Y^2-\str_{Y'}^2|}{2\str_Y\str_{Y'}}  \leq 1, \\
s_1(T) \max\{|A|,|B|,|C|,|D|\} , & \qquad {\rm otherwise} ,
\end{array}
\right.
\end{align}
with $A,B,C,D$ and $J_{\max}$ as defined in equations~(\ref{abcd}) and~(\ref{jmax}).
 Further, the bounds are achieved for the relative angles given by
\begin{align} \label{optang1}
	\cos\theta = 	 \frac{(\str_{X}^2+\str_{X'}^2) (\str_{Y}^2-\str_{Y'}^2)}{2\str_{X}\str_{X'}(\str_{Y}^2+\str_{Y'}^2)}  ,\qquad 
\cos\phi = 	  \frac{(\str_{X}^2-\str_{X'}^2)(\str_{Y}^2+\str_{Y'}^2)}{2\str_{Y}\str_{Y'}(\str_{X}^2+\str_{X'}^2)} ,
\end{align}
when the first condition in equation~(\ref{thm4}) is satisfied, and by
\beq \label{optang2}
\cos\theta={\rm sign}(\str_{Y}-\str_{Y'}),\qquad \cos\phi={\rm sign}(\str_X-\str_{X'})
\eeq
otherwise.
\end{Theorem}

This theorem is proved in~\ref{appc}, where it is also shown that the first form of $S^*_0$ in equation~(\ref{thm4}) is always larger than the second form. Note further that the second form satisfies, taking $\str_X\geq\str_{X'}, \str_Y\geq\str_{Y'}$ without loss of generality, 
\beq
S^*_0+J_{\max} = s_1(T)A+J_{\max}\leq A+J_{\max} = 2 -2\str_{X'}(1-\str_Y) - 2\str_{Y'}(1-\str_X) \leq 2 .
\eeq
Hence, the CHSH inequality can only be violated for states with $s_1(T)=s_2(T)$ if the condition on the first line of equation~(\ref{thm4}) is satisfied.  Further, even when  this condition is satisfied, a violation of the CHSH inequality with unbiased measurements is possible if and only if the strengths further satisfy 
\beq 
(\str_{X}^2+\str_{X'}^2) (\str_{Y}^2+\str_{Y'}^2) > 2/s_1(T)^2\geq 2. 
\eeq 
Thus, Theorem~\ref{theorem4} provides a significant generalisation of the Horodecki criterion~(\ref{horo}) for states with $s_1(T)=s_2(T)$.

Each of Theorems~\ref{theorem3} and~\ref{theorem4} implies that it is possible to violate the CHSH inequality even if one of the measurements is arbitrarily weak. For example, for equal strengths $\str_X=\str_{X'}$, the tight upper bounds in Theorem~\ref{theorem3} allow $\blk S(X,X',Y,Y') \blk>2$ for arbitrarily small (but nonzero) $\str_{Y'}$ and $s_2(T)$, provided that $s_1(T)\str_X\str_Y$ is sufficiently close to 1. Note for the case of arbitrarily small $s_2(T)$ that the state is arbitrarily close to a separable state.  Similarly, for states with $s_1(T)=s_2(T)$, if $\str_X$ and $\str_{X'}$ are sufficiently close such that the condition on the first line of equation~\ref{thm4} of Theorem~\ref{theorem4} is satisfied, then $S^*_0$ and $S^*_T>2$ for arbitrarily small (but nonzero) $\str_{Y'}$, provided that $s_1(T)\str_{Y}\sqrt{\str_X^2+\str_{X'}^2}$ is sufficiently close to $\sqrt{2}$.

 However, if one of the strengths vanishes, $\str_{Y'}=0$ say, then it is impossible to violate the CHSH inequality, for any choice of the other observables and for any state, as might be expected. In particular, in this case $Y'=\B_{Y'}\id$, and so can be measured by simply flipping a suitably biased coin independently of the state, implying that joint probability distributions exist for the triples $(X,Y,Y')$ and $(X',Y,Y')$.  Proposition~1 of Fine~\cite{Fine1982} then immediately implies the existence of a deterministic local hidden variable model for the correlations, and hence that the CHSH inequality cannot be violated. A value of $\blk S(X,X',Y,Y') \blk=2$ is still obtainable, however, even if all strengths vanish, as per the example following equation~(\ref{supper}).

\blk 
\section{Connections between the CHSH inequality and the compatibility of qubit observables}
\label{sec:com}

It was shown by Busch that two unbiased qubit observables $X$ and $X'$ are compatible, i.e., jointly measurable, if and only if they satisfy the condition~\cite{Busch86}
\beq \label{busch}
|\str_X \bm x +\str_{X'} \bm x'| + |\str_X \bm x-\str_{X'} \bm x'|\leq 2.
\eeq 
We note this condition is algebraically equivalent to 
\beq \label{busch2}
\sin\theta\leq \frac{\sqrt{1-\str_{X}^2}}{\str_X} \,\frac{\sqrt{1-\str_{X'}^2}}{\str_{X'}}  .
\eeq
Andersson {\it et al.} have shown that the necessity of Busch's condition for unbiased observables follows from the CHSH inequality~\cite{Andersson05}. Here we generalise their argument  to obtain a necessary condition for arbitrary observables.

First, if $X$ and $X'$ are compatible, then the triples $(X,X',Y)$ and $(X,X',Y')$ each have a classical joint probability distribution, and hence the CHSH inequality~(\ref{chsh}) must be satisfied~\cite{Fine1982}. Second, if the observers share a singlet state, then the substitution of equation~(\ref{sabt}) into~(\ref{chsh}) with $T=-I_3$ gives 
\beq \nn
\str_Y(\str_X \bm x +\str_{X'}\bm x')\cdot \bm y + \str_{Y'}(\str_X \bm x-\str_{X'} \bm x')\cdot \bm y'  + J\leq 2 ,
\eeq
and optimising over the unit directions $\bm y, \bm y'$ and using equation~(\ref{jay}) then yields
\beq \nn
\str_Y|\str_X \bm x +\str_{X'} \bm x'| + \str_{Y'}|\str_X \bm x-\str_{X'} \bm x'| + \B_Y(\B_X+\B_{X'}) + \B_{Y'} (\B_X-\B_{X'})\leq 2.
\eeq
Finally, optimising over $\str_Y,\str_{Y'}, \B_Y,\B_{Y'}$ under the linear constraint~(\ref{sbcon}) gives the simple necessary condition
\beq \label{necbias}
\max\{|\str_X \bm x +\str_{X'} \bm x'|, |\B_X+\B_{X'}|\} + 
\max\{|\str_X \bm x-\str_{X'} \bm x'|, |\B_X - \B_{X'}|\} \leq 2
\eeq
for the compatibility of observables $X$ and $X'$. Note for unbiased observables, $\B_X=\B_{X'}=0$, that this reduces to Busch's  condition~(\ref{busch}), as expected.

The necessary condition~(\ref{necbias}) for the compatibility of $X$ and $X'$ may be compared with the known necessary and sufficient condition~\cite{teiko,busch,oh}, which may be written as~\cite{oh}
\beq \label{suffbias}
(1-\R_X^2-\R_{X'}^2)\, (1-\tfrac{\B_X^2}{\R_X^2} -\tfrac{\B_{X'}^2}{\R_{X'}^2})\leq \left(\str_X\str_{X'} \cos\theta - |\B_X\B_{X'}|\right)^2 ,
\eeq
where
\beq
\R_X := \half\sqrt{(1+\B_X)^2-\str_{X}^2} +  \half\sqrt{(1-\B_X)^2-\str_{X}^2}
\eeq
(and similarly for $\R_{X'}$). 
It is clear that conditions~(\ref{necbias}) and~(\ref{suffbias}) are not equivalent, implying that the former is only necessary but not sufficient. It would be of interest to determine whether the assumption of a singlet state in the derivation of equation~(\ref{necbias}) is responsible for this, and whether (\ref{necbias}) can be improved by optimising $S_T$ in equation~(\ref{thm2}) over $Y$ and $Y'$. It is also of interest to note that the parameter $\R_X$ has been recently interpreted as the `maximum reversibility' of observable $X$~\cite{Cheng21,ChengPRA}.

\blk

\section{Conclusions}
\label{sec5}

We have studied the problem of when a two-qubit state is able to manifest Bell nonlocality via violation of the CHSH inequality, for scenarios in which \blk nonprojective \blk qubit measurements are either desirable or unavoidable. We derived generalised Horodecki criteria for unbiased measurements on arbitrary states and for arbitrary measurements on T-states, that provide necessary and sufficient conditions to be able to violate the CHSH inequality for measurements having fixed strengths and relative angles for each observer (Theorems~\ref{theorem1} and~\ref{theorem2}). We also determined the optimal angles for these cases under the restrictions of equal strengths on one side and/or equal singular values of the spin correlation matrix (Theorems~\ref{theorem3} and~\ref{theorem4}). In all cases the corresponding achievable maximum value of the CHSH parameter was determined.

The results have a number of interesting direct implications, including: 
\begin{itemize}
	\item
	if the measurement strengths on one side are equal, then it is optimal to measure in orthogonal directions on the other side (Theorem~\ref{theorem3}), although this can be relaxed to a one-parameter range of optimal angles for the degenerate case in which the measurement strengths are equal for each side (Corollaries~\ref{corollary1} and~\ref{corollary4});
	\item
	biased measurements may outperform unbiased ones, for a fixed set of measurement strengths, including cases where it is possible to violate the CHSH inequality via biased measurements but not via unbiased measurements (Corollary~\ref{corollary5});
	\item
	in some cases it is possible to violate the CHSH inequality even if one measurement is arbitrarily weak  and/or if the state is arbitrarily close to a separable state (Theorems~\ref{theorem3} and~\ref{theorem4}).
\end{itemize}
The results also extend in a simple way, to give generalised Horodecki criteria valid for all four CHSH inequalities (Corollaries~\ref{corollary3} and~\ref{corollary6}). Further, a strong necessary condition for violation of the CHSH inequality, valid for arbitrary states and measurements, follows from equation~(\ref{sgen}), \blk and we have used the CHSH inequality to obtain a simple necessary condition for the compatibility of generalised qubit observables in section~\ref{sec:com}. \blk

A special case of Theorem~\ref{theorem1}  was 
critical to deriving one-sided monogamy relations for recycling Bell nonlocality in~\cite{Cheng21,ChengPRA}, and we believe our stronger methods and results in this paper may more generally be useful for investigating quantum resources and information protocols in the context of \blk generalised observables. \blk For example, it would be of interest  to investigate a similar approach to bounding violations of other Bell inequalities, such as the Clauser-Horne inequality~\cite{Clauser74} and the Collins-Gisin $I_{3322}$ inequality~\cite{Collins04}, both for two-qubit states and for higher-dimensional entangled states, as well as to apply our methods to the certification of entanglement~\cite{ew} and quantum steering~\cite{steer}.  It would also be of interest to 
sharpen equation~(\ref{sgen}) to obtain a general necessary and sufficient condition, i.e., a full generalisation of the Horodecki criterion to arbitrary states and measurements, and to \blk investigate whether the derivation of the simple compatibility condition~(\ref{necbias}) can be strengthened to obtain the known necessary and sufficient condition~(\ref{suffbias}). \blk  Finally,  the results suggest an experimental investigation of the optimality of biased measurements over unbiased measurements in certain scenarios, as per Corollary~\ref{corollary5}.

{\flushleft{\it Acknowledgements:}
   We thank Travis J. Baker for helpful discussions and comments.  S. C.~is supported by the Fundamental Research Funds for the Central Universities (No.~22120210092) and the National Natural Science Foundation of China (No.~62088101). }

 \appendix

\section{ Proof of Theorems~1 and~2}\label{appa1}
\label{appa}

For general two-valued qubit observables $X,X',Y,Y'$, represented by operators of the form of equation~(\ref{observable}) with $X=\B_X\id+\str_X \bm\sigma\cdot\bm x$, etc., define the unit vectors
\beq \label{xdef}
\bm x_1 = \frac{\bm x+ \bm x'}{|\bm x+\bm x'|},~~~\bm x_2 = \frac{\bm x- \bm x'}{|\bm x - \bm x'|}, ~~~\bm x_3=\bm x_1\times \bm x_2,
\eeq
\beq \label{ydef}
\bm y_1 = \frac{\bm y+ \bm y'}{|\bm y + \bm y'|},~~~\bm y_2 = \frac{\bm y- \bm y'}{|\bm y - \bm y'|}, ~~~\bm y_3=\bm y_1\times \bm y_2.
\eeq
It follows that 
\beq \label{x1x2}
\bm x =\cos\frac{\theta}{2} \bm x_1+\sin \frac{\theta}{2} \bm x_2, ~~~\bm x' =\cos\frac{\theta}{2} \bm x_1-\sin \frac{\theta}{2} \bm x_2,
\eeq
\beq \label{y1y2}
\bm y =\cos\frac{\phi}{2} \bm y_1+\sin \frac{\phi}{2} \bm y_2,~~~ \bm y' =\cos\frac{\phi}{2} \bm y_1-\sin \frac{\phi}{2} \bm y_2 ,
\eeq	
where $\cos \theta=\bm x\cdot \bm x'$ and $\cos \phi=\bm y\cdot\bm y'$, i.e., $0\leq\theta\leq\pi$ is the  angle between $\bm x$ and $\bm x'$ and $0\leq\phi\leq\pi$ is the  angle between $\bm y$ and $\bm y'$. 

The common term in equations~(\ref{chshzero}) and~(\ref{sabt}) in the main text then reduces to	
\begin{align}
&\str_X\str_{Y} \bm x^\top T\bm y + \str_X\str_{Y'} \bm x^\top T\bm y' + \str_{X'}\str_{Y} \bm x'^\top T\bm y - \str_{X'}\str_{Y'} \bm x'^\top T\bm y'\nn\\
	&\qquad\qquad\qquad= 	\sum_{j,k} \tilde W_{jk} \bm x_j^\top T \bm y_k = \tr{\tilde WM^\top},
	\label{wm}
\end{align}
where $\tilde W$ is the $3\times3$-matrix 
\beq
\tilde W:=\begin{pmatrix}
	A\cos\frac{\theta}{2}\cos\frac{\phi}{2} & B\cos\frac{\theta}{2}\sin\frac{\phi}{2} & 0\\ C\sin\frac{\theta}{2}\cos\frac{\phi}{2} & -D\sin\frac{\theta}{2}\sin\frac{\phi}{2} &0 \\ 0& 0 & 0
\end{pmatrix}
\eeq
with $A,B,C,D$ as per equation~(\ref{abcd}), i.e.,
\begin{align}
	A &= \str_{X}\str_{Y}+\str_{X}\str_{Y'}+\str_{X'}\str_{Y}-\str_{X'}\str_{Y'}\nn\\
	B &= \str_{X}\str_{Y}-\str_{X}\str_{Y'}+\str_{X'}\str_{Y}+\str_{X'}\str_{Y'} \nn\\
	C &= \str_{X}\str_{Y}+\str_{X}\str_{Y'}-\str_{X'}\str_{Y}+\str_{X'}\str_{Y'} \nn\\
	D &= -\str_{X}\str_{Y}+\str_{X}\str_{Y'}+\str_{X'}\str_{Y}+\str_{X'}\str_{Y'},
\end{align}
and $M$ is the $3\times3$ matrix with coefficients
\beq
M_{jk} := \bm x_j^\top T \bm y_k .
\eeq
Note that $\tilde W$ has the $2\times2$ matrix $W$ in equation~(\ref{wdef}) as a submatrix and contains local information about the measurement strengths and the relative measurement directions on each side. In contrast, $M$ contains global information about the spin correlation matrix and the relative directions between the two sides.

We then have the following useful result.
{\flushleft \bf Lemma:} {\it For matrices $\tilde W$ and $M$ defined as above, we have the upper bound
	\beq
	|\tr{\tilde WM^\top}|\leq s_1(W)s_1(T) + s_2(W)s_2(T) ,
	\eeq
	where the bound is achievable for any specified measurement strengths and relative angles.}

{\flushleft \it Proof:} First, noting the trace of a matrix is equal to the sum of its eigenvalues, we have
\begin{align}
	\big|\tr{\tilde WM^\top}\big|&=\big|\sum_j\lambda_j(\tilde WM^\top)\big| \leq \sum_j |\lambda_j(\tilde WM^\top)|\nn\\
	&\leq\sum_j s_j(\tilde WM^\top) 
	\leq \sum_j s_j(\tilde W)s_j(M) \nn\\
	&=s_1(W)s_1(T) + s_2(W) s_2(T),
	\label{lemmaproof} 
\end{align}
where $\lambda_j(M)$ denotes the $j$-th eigenvalue of matrix $M$. Here the inequalities on the second line follow from equation~II.23 and Theorem~IV.2.5 of~\cite{Bhatia97}, respectively, and the last line follows from  $s_1(\tilde W)=s_1(W)$, $s_2(\tilde W)=s_2(W)$, $s_3(\tilde W)=0$ (implied by the block-diagonal form of $\tilde W$) and the equality $s_j(M)=s_j(T)$. To prove the latter equality, note that the coordinate systems defined by equations~(\ref{xdef}) and~(\ref{ydef}) are right-handed by construction, implying there must be some rotation matrix $R$ such that $\bm y_k=R\bm x_k$. Hence, 
\beq
M_{jk} = \bm x_j^\top (TR) \bm x_k ,
\eeq
i.e., matrix $M$ is equal to $TR$ with respect to the $\{\bm x_j\}$ basis. But any two matrices differing only by  rotation multipliers have the same singular values~\cite{Bhatia97}, yielding the desired result.  

Second, to show that the upper bound of the Lemma is achievable, note that applying arbitrary orthogonal transformations $O_1$ and $O_2$ to the $\{\bm x_j\}$ and $\{\bm y_j\}$ bases, respectively, keeps the strengths and relative angles $\theta,\phi$, and hence the matrix $\tilde W$, invariant, while changing the coefficients of $M$ to
\beq
M'_{jk} =\bm (O_1\bm x_j)^\top T O_2\bm y_k = \bm x_j^\top (O_1^\top TO_2R) \bm x_k,
\eeq
where $R$ is the rotation matrix defined above. Thus, $M'$ is equal to $O_1^\top TO_2R$ with respect to the $\bm x_j$ basis, and it follows via equations~(\ref{xdef}) and~(\ref{ydef}) that the mapping from $M$ to $M'$ corresponds to transforming the measurement directions $\bm x,\bm x'$ by $O_1$ and the measurement directions $\bm y,\bm y'$ by $O_2$, while keeping all other properties of the observables invariant. It further follows that
\beq
\tr{\tilde WM'^\top}=\tr{\tilde WO_1^\top TO_2R} .
\eeq
Now, let $\tilde W=P_1D(\tilde W)P_2^\top$ and $T=Q_1D(T)Q_2^\top$ be singular value decompositions of $\tilde W$ and $T$, for suitable orthogonal matrices $P_1,P_2,Q_1,Q_2$, where $D(\tilde W)$ and $D(T)$ denote the diagonal matrices with coefficients $D(\tilde W)_{jj}=s_j(\tilde W), D(T)_{jj}=s_j(T)$. Finally, using the cyclicity of the trace and choosing 
\beq \label{o1o2}
O_1=P_2Q_1^\top, \qquad O_2=Q_2 P_1^\top R^\top  
\eeq  
gives
\begin{align}
	\tr{\tilde WM'^\top}&=\tr{D(\tilde W) P_2^\top O_1^\top Q_1D(T)Q_2^\top O_2R  P_1} \nn\\
	&=\tr{D(\tilde W)D(T)} = \sum_j s_j(\tilde W) s_j(T)\nn\\
	&=  s_1(W)s_1(T) + s_2(W)s_2(T)
\end{align}
as desired, where the last line follows via the previously noted relations  $s_1(\tilde W)=s_1(W)$, $s_2(\tilde W)=s_2(W)$, $s_3(\tilde W)=0$. $\blacksquare$

The above Lemma, together with equations~(\ref{chshzero}) and~(\ref{wm}), immediately implies the tight upper bound $S_0$ for unbiased observables in equation~(\ref{sabzero}) of Theorem~\ref{theorem1}. Note that the optimal measurement directions are determined from some reference set of directions $\bm x,\bm x', \bm y, \bm y'$, having relative angles $\theta$ and $\phi$, by applying the orthogonal transformations $O_1$ and $O_2$ in equation~(\ref{o1o2})to $\bm x,\bm x'$ and $\bm y,\bm y'$, respectively.

The tight upper bound $S_T$ for T-states in equation~(\ref{thm2}) of Theorem~\ref{theorem2}  similarly follows, provided it can be shown that the maximum value of $|J|$ in equation~(\ref{jay}) is given by $J_{\max}$ in equation~(\ref{jmax}). 
Indeed, since the biases contribute linearly to $J$, and $|\B|\leq\bar\str:= 1-\str$ as per equation~(\ref{sbcon}), it follows that the maximum and minimum values of  $J$ correspond to $\B_X=\alpha\bar\str_X$, $\B_{X'}=\alpha'\bar\str_{X'}$, $\B_Y=\beta\bar\str_{Y}$, $\B_{Y'}=\beta'\bar\str_{Y'}$, for suitable choices of $\alpha_,\alpha',\beta,\beta'=\pm1$.  
We then have, defining $\gamma=\beta\beta'$,
\begin{align}
	|J|&\leq\max_{\alpha,\alpha',\beta,\beta'=\pm1} |\alpha\bar\str_X(\beta\bar\str_Y+\beta'\bar\str_{Y'}) + \alpha'\bar\str_{X'}(\beta\bar\str_Y-\beta'\bar\str_{Y'})| \nn\\
	&= \max_{\gamma=\pm1}\, \bar\str_{X}|\bar\str_Y+\gamma\bar\str_{Y'}| + \bar\str_{X'}|\bar\str_Y-\gamma\bar\str_{Y'}| \nn\\
	&=\max\{\bar\str_X,\bar\str_{X'}\}(\bar\str_{Y}+\bar\str_{Y'}) + \min\{\bar\str_X,\bar\str_{X'}\}|\bar\str_{Y}-\bar\str_{Y'}|\nn\\
	&= (\bar\str_X+\bar\str_{X'})(\bar\str_{Y}+\bar\str_{Y'}) -2\min\{\bar\str_X,\bar\str_{X'}\}\min\{\bar\str_Y,\bar\str_{Y'}\}\nn\\
	&= (2-\str_X-\str_{X'})(2-\str_Y-\str_{Y'})\nn\\
	&\qquad - 2(1-\max\{\str_X,\str_{X'}\})(1-\max\{\str_Y,\str_{Y'}\})\nn\\
	&= J_{\max}.
	\label{proofthm2}
\end{align}
Here the third line follows using $ac+bd\geq ad+bc$ for $a\geq b, c\geq d$ and the fourth line by verifying it for the case $\bar S_X\leq\bar S_{X'}, \bar S_Y\leq\bar S_{Y'}$ and then applying the interchange symmetries $X\leftrightarrow X'$, $Y\leftrightarrow Y'$ of the third line. Note that the upper bound is tight, being achieved for a given value of $\beta=\pm1$ by the choices $\beta'=\beta\,{\rm sign}(\bar\str_{X}-\bar\str_{X'})$, $\gamma=\beta\beta'$, $\alpha={\rm sign}(\beta\bar\str_{Y}+\beta'\bar\str_{Y'})$ and $\alpha'={\rm sign}(\beta\bar\str_{Y}-\beta'\bar\str_{Y'})$. Hence, Theorem~\ref{theorem2} of the main text follows as desired. 

Finally, to obtain the alternative expression for $S_0$ in equation~(\ref{cor2}) of Theorem~\ref{theorem1}, note first that $S_0$ in equation~(\ref{sabzero}) can be rewritten as
\begin{align}
	S_0 =\half I_+(W)[s_1(T)+s_2(T)]+ \half I_-(W)[s_1(T) - s_2(T)] ,
\end{align}
with $I_\pm(W)=s_1(W)\pm s_2(W)=\sqrt{w_+}\pm \sqrt{w_-}$, where $w_\pm$ are the eigenvalues of $W^\top W$. 
The identities
\beq 
w_+ + w_- = \tr{W^\top W},~ w_+w_-=\det(W^\top W)=\det(W)^2, \nn
\eeq
then imply that
\begin{align}
	I_\pm(W)^2&=
	(\sqrt{w_+}\pm\sqrt{w_-})^2 =
	w_++w_- \pm 2\sqrt{w_+w_-} = \tr{W^\top W} \pm 2 |\det(W)| .
	\label{cor2proof}
\end{align}
Explicit calculation of $\tr{W^\top W}$ and $\det(W)$ from the definitions in equations~(\ref{wdef}) and~(\ref{abcd}) yields equation~(\ref{iw}) for $I_\pm(W)^2$, and equation~(\ref{cor2}) follows as desired.

\section{Proof of Corollaries 1 and 4}
\label{appa2}

To obtain the tight bound in Corollary~\ref{corollary1} of the main text,  note for the case of equal strengths on each side that the eigenvalues $w_\pm$ of $W^\top W$ simplify  to
\beq \label{wpmequal}
w_\pm = 2\str_X^2\str_Y^2\left(1\pm \sqrt{1-\sin^2 \theta\sin^2 \phi}\right) .
\eeq
Using $s_1(W)=w_+^{1/2}, s_2(W)=w_-^{1/2}$ and applying the Schwarz inequality to equation~(\ref{sabzero}) of Theorem~\ref{theorem1} then gives
\begin{align} 
	\blk S(X,X',Y,Y') \blk &\leq \sqrt{w_++w_-}\sqrt{s_1(T)^2+s_2(T)^2}\nn\\
	&=2\str_X\str_Y\sqrt{s_1(T)^2+s_2(T)^2},
	\label{cor1proof}
\end{align}
with equality if and only if $w_-/w_+=s_2(T)^2/s_1(T)^2$. But, as easily verified by direct substitution into equation~(\ref{wpmequal}), the latter condition is equivalent to choosing the relative angles $\theta$ and $\phi$ as per equation~(\ref{optang}), yielding Corollary~\ref{corollary1} as desired. 
The proof of Corollary~\ref{corollary4} is entirely analogous,  where the use of Theorem~\ref{theorem2} in place of Theorem~\ref{theorem1} leads to the addition of $J_{\max}$ to the right hand side of equation~(\ref{cor1proof}).

\section{Proof of Theorems~3 and~4} \label{appc}

 The upper bounds in Theorem~\ref{theorem3} may be obtained in several ways. For example, for $\str_X=\str_{X'}$ the matrix $W$ in equation~(\ref{wdef}) simplifies to $W=LW_0$, where $L={\rm diag}[\str_Y,\str_{Y'}]$ and $W_0$ is given by replacing $\str_Y$ and~$\str_{Y'}$ by 1 in~(\ref{wdef}), which allows the techniques in~\ref{appa} to be generalised to upper bound $\blk S(X,X',Y,Y') \blk=|\tr{W_0LM^\top}|$. 
However, here we will instead directly upper bound the CHSH parameters in equations~(\ref{chshzero}) and~(\ref{sabt}), corresponding to the two cases considered in Theorem~\ref{theorem3}.

First, the equal strength assumption $\str_X=\str_{X'}$ reduces equation~(\ref{chshzero}) for unbiased observables to
\begin{align}
	\blk S(X,X',Y,Y') \blk &=\str_{X}\left|(\bm x+\bm x^\prime)^\top T (\str_{Y}\bm y)+(\bm x-\bm x^\prime)^\top T (\str_{Y'}\bm y')\right| \nn \\
	&=2\str_{X}\left|\cos\frac{\theta}{2} \bm x_1^\top T (\str_{Y}\bm y)+\sin\frac{\theta}{2}\bm x_2^\top T (\str_{Y'}\bm y')\right| \nn \\
	&=2\str_{X}\left|\cos\frac{\theta}{2}\str_Y \bm y\cdot (T^\top\bm x_1) +\sin\frac{\theta}{2}\str_{Y'}\bm y'\cdot(T^\top\bm x_2) \right| \nn \\
	&\leq 2\str_{X}\left[\cos\frac{\theta}{2} \str_{Y}|T^\top \bm x_1| +\sin\frac{\theta}{2}\str_{Y'}|T^\top \bm x_2|\right] \nn \\
	&\leq 2\str_{X}\rt{\str^2_{Y}|T^\top \bm x_1|^2+\str_{Y'}^2|T^\top \bm x_2|^2} \nn\\ 
	&=2\str_X\rt{\tr{TT^\top K}}	, \label{thm3proof}
\end{align}
where the orthogonal unit directions $\bm x_1, \bm x_2$ are defined in equation~(\ref{xdef}), the first and second inequalities are saturated by choosing
\beq \label{sat}
\bm y = T^\top \bm x_1/|T^\top \bm x_1|, \qquad \bm y' = T^\top \bm x_2/|T^\top \bm x_2|, \qquad
\tan\frac{\theta}{2} = \frac{\str_{Y'}|T^\top \bm x_2|}{\str_{Y}|T^\top \bm x_1|} ,
\eeq
and $K$ denotes the $3\times3$ matrix defined by
\beq 
K:=S_Y^2\bm x_1\bm x_1^\top + S_{Y'}^2\bm x_2\bm x_2^\top .
\eeq
It then follows from equations~(\ref{lemmaproof}) and(\ref{thm3proof}), for $\str_Y\geq\str_{Y'}$, that
\begin{align} \label{upper}
\blk S(X,X',Y,Y') \blk&\leq 2\str_X\left[\sum_j s_j(TT^\top) s_j(K)  \right]^{1/2}
= 2\str_{X}\rt{\str_{Y}^2s_1(T)^2+\str_{Y'}^2 s_2(T)^2}
\end{align}
as per the upper bound in equation~(\ref{thm30}) of Theorem~\ref{theorem3}, with equality achieved by choosing the orthogonal directions $\bm x_1, \bm x_2$ to be the eigenvectors of $TT^\top$ corresponding to eigenvalues $s_1(T)^2, s_2(T)^2$, respectively. Further, substituting this choice into equation~(\ref{sat}) above yields
\beq
\cos\phi = \bm y\cdot\bm y' = \frac{\bm x_1^T TT^\top \bm x_2}{|T^\top \bm x_1||T^\top \bm x_2|}=0, ~~~ \tan\frac{\theta}{2} = \frac{\str_{Y'}\sqrt{\bm x_2^\top TT^\top \bm x_2}}{\str_{Y}\sqrt{\bm x_1^\top TT^\top \bm x_1}} =  \frac{\str_{Y'}s_2(T)}{\str_{Y}s_1(T)} ,
\eeq
which is equivalent to equation~(\ref{optangone}) of Theorem~\ref{theorem3}. Note that for the degenerate case $\str_{Y}=\str_{Y'}$ equality can also be achieved in equation~(\ref{upper}) for any orthogonal directions $\bm x_1, \bm x_2$ that lie in the span of the eigenvectors of $TT^\top$ corresponding to eigenvalues $s_1(T)^2, s_2(T)^2$. This freedom can be shown to be equivalent to the condition~(\ref{optang}) of Corollary~\ref{corollary1}, as expected. Finally, the upper bound in equation~(\ref{thm3t}) for arbitrary measurements on a T-state follows by applying the same analysis to the CHSH parameter in equation~(\ref{sabt}), and substituting $\str_X=\str_{X'}$ and $\str_Y\geq \str_{Y'}$ into equation~(\ref{jmax}) for $J_{\max}$.

To obtain the upper bounds in Theorem~\ref{theorem4} note first that, under the condition $s_1(T)=s_2(T)$ of the theorem, the tight bounds in Theorems~\ref{theorem1} and~\ref{theorem2} simplify to
\beq \label{boundsimp}
S_0 = s_1(T) I_+(W), \qquad S_T = s_1(T) I_+(W) + J_{\max},
\eeq
where $I_+(W)$ is defined in equation~(\ref{iw}). Further, from the latter equation we have 
\beq \label{iwsimp}
I_+(W)^2 = 	(\str_{X}^2+\str_{X'}^2)(\str_{Y}^2+\str_{Y'}^2) + 2f(\theta,\phi),
\eeq
with $f(\theta,\phi):=a\cos\theta+b\cos\phi + c\sin\theta\sin\phi$ and
\beq \label{abc}
a:=\str_X\str_{X'}(\str_{Y}^2-\str_{Y'}^2),~~b:=\str_Y\str_{Y'}(\str_{X}^2-\str_{X'}^2),~~c:=
2\str_{X}\str_{X'}\str_{Y}\str_{Y'}  .
\eeq
Setting the partial derivatives of $f$ with respect to $\theta$ and $\phi$ equal to zero gives
\beq \label{deriv}
a\sin\theta=c\cos\theta\sin\phi, \qquad b\sin\phi=c\sin\theta\cos\phi .
\eeq
Further, multiplying these equations together yields $\sin\theta\sin\phi(ab-c^2\cos\theta\cos\phi)=0$, implying that
\beq
 \sin\theta=0 ~~{\rm or}~~ \sin\phi=0 ~~ {\rm or}~~ \cos\theta\cos\phi=\frac{ab}{c^2}.
\eeq
Maximising $f$ under either of the first two constraints gives a first solution
\beq \label{sol1}
\cos\theta_1 = {\rm sign}(a),\qquad \cos\phi_1 = {\rm sign}(b),~~f_1=|a|+|b| ,
\eeq
while the third constraint can yield a solution if and only if the consistency condition
\beq
1\geq |\cos\theta\cos\phi|=\frac{|ab|}{c^2} = \frac{|\str_X^2-\str_{X'}^2|}{2\str_X\str_{X'}}  \frac{|\str_Y^2-\str_{Y'}^2|}{2\str_Y\str_{Y'}} 
\eeq
is satisfied, in which case it is straightforward to check that equation~(\ref{deriv}) yields a second maximum, with
\beq \label{sol2}
\tan\theta_2 = \frac{\sqrt{c^4-a^2b^2}}{a{\sqrt{b^2+c^2}}}, ~~\tan\phi_2 = \frac{\sqrt{c^4-a^2b^2}}{b{\sqrt{a^2+c^2}}}, ~~ f_2=  (1/c)\sqrt{(a^2+c^2)(b^2+c^2)} .
\eeq
It follows that
\beq
f_2^2-f_1^2  = |ab|\left(\frac{|ab|}{c^2}+\frac{c^2}{|ab|}\right) -2|ab| \geq 2|ab|-2|ab|= 0,
\eeq
and hence the second solution, when it exists, is the global maximum.

Substitution of equation~(\ref{abc}) into equations~(\ref{sol1}) and~(\ref{sol2}) yields the  optimal angles in equations~(\ref{optang1}) and~(\ref{optang2}) of Theorem~\ref{theorem4} (using $\cos^2x=1/(1+\tan^2x)$ and noting $\sin\theta,\sin\phi\geq0$), and the explicit expressions
\beq
f_1 = \str_X\str_{X'}|\str_{Y}^2-\str_{Y'}^2| + \str_Y\str_{Y'}|\str_{X}^2-\str_{X'}^2| ,
~~
f_2 = \half (\str_X^2+\str_{X'}^2)(\str_Y^2+\str_{Y'}^2) .
\eeq
Finally, inserting these expressions into equations~(\ref{boundsimp}) and~(\ref{iwsimp}) yields the upper bounds in equations~(\ref{thm4upper})~and(\ref{thm4}) of Theorem~\ref{theorem4}, as desired (for $f_1$ it is easiest to first assume $\str_X\geq\str_{X'}$ and $\str_Y\geq \str_{Y'}$, to obtain $S^*_0=|A|$, and then generalise).


~\\

\end{document}